\documentclass[conference]{IEEEtran}

\usepackage{amsmath,amssymb,latexsym,hhline,graphicx,enumerate,cite,subfigure}
\usepackage{psfrag, color}
\usepackage{authblk}
\usepackage{cite}
\usepackage[margin=15mm]{geometry}

\usepackage{tikz}
\usetikzlibrary{calc}

\makeatletter
\def\ps@headings{%
\def\@oddhead{\mbox{}\scriptsize\rightmark \hfil \thepage}%
\def\@evenhead{\scriptsize\thepage \hfil \leftmark\mbox{}}%
\def\@oddfoot{}%
\def\@evenfoot{}}
\makeatother
\renewcommand{\cal}{\mathcal}
\newcommand{\nchoosek}[2]{\binom{{#1}}{{#2}}}
\newcommand{\mn}{\mathrm{n}}
\newcommand{\mumax}{\mu_{\mathrm{s}}(\alpha)}
\newcommand{\Pk}{P(k,\alpha)}
\newcommand{\muk}{\mu_{\mathrm{s}}(\alpha|k)}
\newcommand{\Ts}{T_{\mathrm{s}}(\alpha|k)}
\newcommand{\Ps}{P_{\mathrm{s}}}
\newcommand{\Tss}{T_{\mathrm{s}}}

\newtheorem{theo}{Theorem}

\newtheorem{lem}{Lemma}
\newtheorem{corol}{Corollary}

\IEEEoverridecommandlockouts

\begin{document}

% paper title
\title{On Storage Allocation for Maximum Service Rate in Distributed Storage Systems}

\author{
    Moslem Noori\IEEEauthorrefmark{1}, Emina Soljanin\IEEEauthorrefmark{2}, Masoud Ardakani\IEEEauthorrefmark{1}\\
    \IEEEauthorrefmark{1}Department of Electrical and Computer Engineering, University of Alberta, Edmonton, AB, Canada\\
    \IEEEauthorrefmark{2} Department of Electrical and Computer Engineering, Rutgers University, Piscataway, NJ, USA
}

\maketitle

\pagenumbering{gobble}

\begin{abstract}
Storage allocation affects important performance measures of distributed storage systems. Most previous studies on the storage allocation consider its effect separately either on the success of the data recovery or on the service rate (time) where it is assumed that no access failure happens in the system. In this paper, we go one step further and incorporate the access model and the success of data recovery into the service rate analysis. In particular, we focus on quasi-uniform storage allocation and provide a service rate analysis for both fixed-size and probabilistic access models at the nodes. Using this analysis, we then show that for the case of exponential waiting time distribution at individuals storage nodes,  minimal spreading allocation results in the highest system service rate for both  access models. This means that for a given storage budget, replication provides a better service rate than a coded storage solution.  
\end{abstract}

\section{Introduction}
Cloud networks provide anywhere, anytime access to one's data, offer a high level of data safety (e.g., against hardware failure, theft, fire), and make sharing data easy. 
This functionality is achieved by storing chunks of a data entity (file) redundantly over multiple  storage nodes. Distributed storage systems (DSSs), thus play a central role in cloud networks, and have been the focus of many ongoing diverse research activities 
\cite{dimakis2010network, joshi2012coding, chen2014queueing,tandon2014new,kadhe15availability}. 

A main concern for the consumers is to be able to download the data and, often more importantly, to do that quickly. Thus, the download service rate is the focus of this paper. Several studies have looked into how to allocate redundant chunks of data over the storage nodes to optimize some performance metrics (e.g., \cite{leong2012distributed,Sardari_Allocation_2010,noori2015allocation,hong2014asymptotic, leong2011distributed} and references therein). The constraints here are that the number of nodes and the level of redundancy are limited, and to download his file, the user can access all or some subset of (possibly unavailable) nodes in the system. 

Existing studies on the storage allocation mostly focus on two performance aspects of DSSs. One of them is the probability of successful data recovery $\Ps$ when only a subset of possibly failed nodes are  accessed. The other is the average service time $\Tss$ when a set of nodes from which the file can be recovered is accessed. Simply put, when a subset of storage nodes are assigned to serve a customer, $\Ps$ is the probability that these nodes jointly (under possible failures) have been allocated sufficient data to reconstruct and deliver the requested file to the customer. On the other hand, $\Tss$ represents the time needed to serve a customer's request to download the file. In other words, $\Ps$ reflects the reliability of the DSS in serving the customers' requests while $\Tss$ mostly represents the system's quality of service once the reliability has been provided. Finding these quantities has shown to be quite challenging, and optimal allocations are known only in some special cases. 

In general, both these measures are of interest and should be simultaneously taken into account for devising the  allocation strategy. For instance, assume a situation where several customers send a delay-sensitive request to access the stored data. While increasing the chance of successfully downloading the file by each of the customers is desirable, this should not come at the cost of unbearable delivery delay. Moreover, in practice, we may often want to partially sacrifice a successful (but possibly tardy) data delivery to some users in order to ensure that other users, that can receive the data, are indeed served fast. 

The existing work does not address such scenarios. Papers concerned with $\Ps$ are not concerned with the delay or assume instantaneous (infinite rate) service . On the other hand, papers  concerned with $\Tss$ assume that data is available on the accessed nodes and can be served to the customer at some finite rate. 

In this work, we assume a finite service rate for storage nodes and the data (un)availability that depends on the used allocation scheme. We are interested in the entire system service rate, under certain access and/or node failure models. Note that, depending on the allocation, some subsets of nodes will not contain enough file chunks between them to recover the data, and accessing them will result in a zero system's service  rate. On the other hand, again depending on the allocation, some subsets of nodes will contain redundant file chunks, and that redundancy can be exploited to increase the service rate.

Our analysis reveals that the allocation that maximizes the probability of successful data recovery is often not the one that maximizes the average service rate.  The key to understanding this, perhaps unintuitive, phenomenon is to look into the role of redundancy. When the accessed nodes contain more data than necessary to reconstruct the file, this redundancy is superfluous for file recovery but could be exploited to speed up the download service rate since only a fraction of nodes have to deliver their chunks in a timely manner. Therefore, depending on the number of storage nodes and the allocated redundancy budget, it may be beneficial for recovery to maximally spread the redundant file chunks
over the storage nodes, whereas concentrating the redundant chunks may increase the expected service rate. We show here that this is always the case for the DSS models considered in the literature. 

The rest of the paper is organized as follows. In Section~\ref{Sec Model}, we introduce the considered DSS setup in more detail and formally define the considered problem in this paper. Service rate analysis considering the effect of access model and the success of serving a request is presented in Section~\ref{Sec mu analysis}. Using this analysis, we then prove that minimal spreading maximizes the service rate of the system in Section~\ref{Sec Fixed} and Section~\ref{Sec Prob} respectively for the fixed-size and probabilistic access models. Numerical examples are also provided in these two chapters. Finally, Section~\ref{Sec Conc} concludes the paper. 

\section{System Model and Problem Definition} \label{Sec Model}
In this section, we describe the considered DSS in detail. Then, we formally define the storage allocation problem to maximize the service rate of the system.

\subsection{Storage Model}
We consider a DSS with $N$ storage nodes, namely $\mn_i$'s for $i \in \cal{N} = \{ 1,\ldots,N \}$. A file with $F$ blocks is stored over these nodes that are to be accessed by the system's customers. To protect the data against nodes' failure, the file is encoded by a maximum distance separable (MDS) code to generate $T$ encoded blocks (any $F$ of them are sufficient to recover the original file). Here, we assume that the code rate is $1/m$, where $m$ is a positive integer. Hence, $T = m F$. The encoded $T$ blocks are then partitioned into $N$ subsets, say $\cal{X}_i$'s for $i  \in \cal{N}$ where $\vert \cal{X}_i  \vert = x_i$, and thus $\sum_{i = 1}^N x_i = T$.   We call such partitioning an \emph{allocation}. Now, the $x_i$ blocks within $\cal{X}_i$ are stored at the node $\mn_i$. Note that $0 \leq x_i \leq F$ since storing more than $F$ blocks on a node does not serve any purpose in our model. 
 
\begin{figure}%
\psfrag{1}{$1$}
\psfrag{2}{$2$}
\psfrag{ma}{$m \alpha$}
\psfrag{ma1}{$m \alpha + 1$}
\psfrag{ma2}{$m \alpha + 2$}
\psfrag{N}{$N$}
\psfrag{A}{$\cal{A} = \{ 1, 2 , N \}$}
\begin{centering}
\includegraphics[width=0.8\columnwidth]{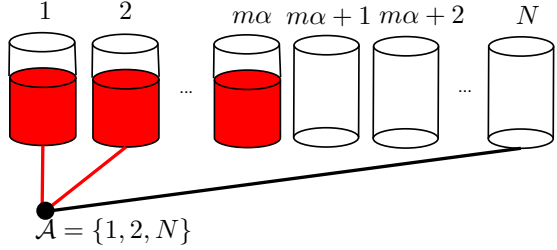}%
\caption{An $N$-node DSS with quasi-symmetric allocation. While three nodes are successfully accessed, only two of them have (coded) data blocks.}%
\label{Fig Model}%
\end{centering}
\end{figure}

Dealing with a general storage allocation optimization problem to maximize $\Ps$ or minimize $\Tss$ is computationally difficult for a general setup \cite{leong2012distributed}. Here, we focus on the quasi-symmetric  allocations \cite{Sardari_Allocation_2010} where for a positive integer $\alpha$, the number of blocks stored in $\mn_i$, denoted by $x_{\alpha}(i)$, is either $0$ or $F/\alpha$. Details of the range of $\alpha$ will be discussed later. Here, we identify a quasi-symmetric allocation with a pair $(\alpha, \beta)$ where $\beta$ represents the number of nodes that are not empty. Since $\beta \frac{F}{\alpha} = T$, we have $\beta = m \alpha$. Figure~\ref{Fig Model} depicts an example quasi-symmetric allocation for a DSS with $N$ storage nodes.

A quasi-symmetric allocation where $\alpha = 1$ and $\beta = m$ is called a \emph{minimal spreading} allocation \cite{leong2012distributed}. Note that for a minimal spreading allocation, we can skip coding and replicate the whole $F$ blocks of the file over $m$ storage nodes without compromising the file protection. Similarly, an allocation with $\alpha = \frac{NF}{T}$ and $\beta = N$ is called a \emph{maximal spreading} allocation.

\subsection{Data Access Model}
For the data access by the users, we consider the following two main models suggested for DSSs\cite{leong2012distributed}.  

\subsubsection{Fixed-size access}
In this access model, when a download request is received, the request is forwarded to a random $r$-subset of the $N$ nodes, i.e., a subset with cardinality $r$  \cite{leong2012distributed, Sardari_Allocation_2010}. Since an MDS code is used to store the data, the original file can be recovered if the accessed nodes contain at least $F$ blocks. In other words, the access to a given $r$-subset $\cal{A}$ results in the successful recovery of the data iff 
\begin{equation}\label{eq: Recovery cond}
\sum_{i \in \cal{A}} x_i \geq F.
\end{equation}
Note that for $\alpha > r$, it is impossible to recover the data. Thus, for the fixed-size access model we only consider $1 \leq \alpha \leq r$. 

\subsubsection{Probabilistic access}
In this case, the download request is forwarded to all nodes that store the data. However, the request to access each of them fails with probability $p$ and succeeds with probability $1 - p$. Assuming that $\cal{A}$ represents the set of nodes that are successfully accessed, then the condition for data recovery is similar to (\ref{eq: Recovery cond}). In this case, $1 \leq \alpha \leq \frac{NF}{T}$.

Regardless of the access model, for an arbitrary accessed subset of nodes $\cal{A}$, let us denote the number of nodes containing data by $k$. For instance, in Figure~\ref{Fig Model}, three nodes 
($|\cal{A}|=3$) are accessed while only $k = 2$ of them have data. For an $(\alpha,\beta)$ quasi-symmetric allocation, data recovery from this subset is successful if and only if $k \geq \alpha$.

\subsection{Service Model}
Here, we assume that the arriving download requests follow a Poisson distribution. Each request is forwarded to a set of accessed nodes, called $\cal{A}$, to be served. At these nodes, we assume a multiple-fountain system \cite{joshi2012coding} where the arriving request is forked into $|\cal{A}|$ tasks\footnote{For the fixed-size access model $|\cal{A}| = r$ while for the probabilistic access $|\cal{A}|$ could be any number between 1 and $N$.}. Each of these tasks then wait to be served by one of the accessed $|\cal{A}|$ nodes. For an $(\alpha,\beta)$ allocation over the nodes, the download request is successfully served when $k \geq \alpha$ and any $\alpha$ out of the $k$ nodes with data successfully serve their assigned tasks. At this point, the remaining $|\cal{A}| - \alpha$ tasks are discarded and dropped from the rest of the accessed nodes. If the nodes in $\cal{A}$ do not contain enough data to reconstruct the file (i.e. less than $F$ blocks), the download request cannot be served. 

When a task is assigned to a storage node, it may not get served right away since the node is for example busy with serving another request. Thus, there is a waiting time associated with the time needed for the content inside the node to become available for download by the user. Here, for simplicity, we assume that the waiting time at all nodes are independent and identically distributed (i.i.d) random variables all following an exponential distribution with mean $\frac{1}{\mu}$. In other words, each storage node has a service rate of $\mu$. Further, it is assumed that the download bandwidth is large enough so that the time needed to download the data is negligible compared to the waiting time at the servers. As a result, the overall service rate of the system is characterized only by the waiting time at the servers.

\subsection{Problem Definition}
For a given $(\alpha,\beta)$ quasi-symmetric allocation, the average service rate, denoted by $\mumax$ is the highest rate that the arriving download requests can be served by the system. 
(Since $\beta = m \alpha$, we do not consider $\beta$ as a separate variable here.)
As we discuss  later in the paper, beside the service rate at each individual node, $\mumax$ also depends on the nodes' storage allocation. Our goal in this paper is to find the allocation that maximizes $\mumax$. 

For a formal problem definition, we introduce the function
\[
\mathbb{I}\Bigl\{\sum_{i\in \mathcal{A}}x_{\alpha}(i)\ge F\Bigr\}
\]
indicating whether the file can be recovered from the nodes in $\mathcal{A}$ or not.
The probability of being able to successfully recover the file under an $(\alpha, \beta)$ allocation is, therefore, given by
\begin{equation}
\Ps(\alpha) = 
\sum_{\mathcal{A}\subseteq \mathcal{N}}
P(\cal{A}) \mathbb{I}\Bigl\{\sum_{i\in \mathcal{A}}x_{\alpha}(i)\ge F\Bigr\} 
\label{eq:prob}
\end{equation}
where $P(\cal{A})$ is the probability of choosing $\cal{A}$.
Similarly, the average service rate  under an $(\alpha, \beta)$ allocation  is given by
\begin{equation}
\mumax = 
\sum_{\mathcal{A}\subseteq \mathcal{N}}
P(\cal{A})  \mu_{\alpha}(\mathcal{A}) \mathbb{I}\Bigl\{\sum_{i\in \mathcal{A}}x_{\alpha}(i)\ge F\Bigr\}
\label{eq:service}
\end{equation}
where $\mu_{\alpha}(\mathcal{A})$ is the service rate when the set of accessed nodes is $\mathcal{A}$. 

Previous studies on finding the optimal storage allocations are focused on finding the allocation that maximizes $\Ps$ for a given storage budget $T$. For instance,  it was shown in \cite{Sardari_Allocation_2010} that for a DSS with fixed-size access model, $\alpha$ that maximizes \eqref{eq:prob}  depends on the ratio $m=T/F$. Similar claims are made in \cite{leong2012distributed}. It is easy to see that $\mu_{\alpha}(\mathcal{A})$ is a decreasing function of $\alpha$. Therefore, when $\alpha=1$ maximizes \eqref{eq:prob}, i.e. minimal spreading maximizes $\Ps$,  it also maximizes \eqref{eq:service}, term by term and thus maximizes $\mumax$. We devote the following sections to showing that \eqref{eq:service} is maximized by $\alpha=1$ even if, for some $\alpha>1$, there are more sets $\mathcal{A}\subset \mathcal{N}$ that allow file reconstruction (more non-zero terms in \eqref{eq:service}) than for $\alpha=1$. That is,  the service rate is always maximized by using minimal spreading allocation.

%More formally, we focus on solving the following optimization problem
%\begin{align} \label{eq: optimization}
%\max_{(\alpha)} \, & \mumax \\ \nonumber
%& m \alpha = \beta, \\ \nonumber
%& 1 \leq \alpha \leq r.
%\end{align}

\section{Analysis of $\mumax$} \label{Sec mu analysis}
In this section, we study $\mumax$ considering the effect of storage allocation and access model. This study will then be used in the following sections to find the optimal allocation maximizing $\mumax$ for fixed-size and probabilistic access models. 

The rate of serving incoming requests depends on how many nodes with data are successfully accessed. Thus,
\begin{align}
\mumax = \sum_{k=1}^{m \alpha} \Pk \muk.
\end{align}
where $\Pk$ denotes the probability of having exactly $k$ nodes with data in the set of accessed nodes $\cal{A}$. Also, $\muk$ refers to the conditional service rate given that $k$ nodes with data are accessed. Note that for any $k < \alpha$, recovering the data from the nodes in $\cal{A}$ is not possible, and $\muk = 0$. Thus, 
\begin{equation}\label{eq: mumax def}
\mumax = \sum_{k=\alpha}^{m \alpha} \Pk \muk.
\end{equation}
It is easy to show that for the fixed-size access model
\begin{equation}
\Pk = \frac{\nchoosek{m\alpha}{k} \nchoosek{N - m \alpha}{r - k}}{\nchoosek{N}{r}}
\label{eq: Pk}
\end{equation}
and for the probabilistic access model 
\begin{equation}
\Pk = \nchoosek{m\alpha}{k} (1 - p)^k p^{m\alpha - k}.
\label{eq:Pk prob}
\end{equation}
Now that we have $\Pk$, to evaluate $\mumax$, we present the following result on $\muk$. 

\begin{lem} \label{Lemma mumax} 
For a given $(\alpha,\beta)$ quasi-symmetric allocation, 
\begin{equation}
\muk = \mu \, \frac{\displaystyle{\prod_{j = k - \alpha + 1}^k j}}{\displaystyle{\sum_{i = k - \alpha + 1}^k} \prod_{j = k - \alpha + 1 \atop j \neq i}^k j}.
\label{eq: muk}
\end{equation}
\end{lem}

\begin{IEEEproof}
To find $\muk$, we start by considering the conditional service time of the requests, denoted by $\Ts$, which is the inverse of $\muk$. As discussed before, a request is served when the first $\alpha$ storage nodes with data, out of the accessed $k$ nodes with data, start serving the request. That said, $\Ts$ is the $\alpha$th order statistics of $k$ waiting times at the storage nodes. Considering that all waiting times have an exponential distribution with mean $\frac{1}{\mu}$, we have 
\begin{equation}
\Ts = \frac{1}{\mu} \sum_{i  = 1}^{\alpha} \frac{1}{k - \alpha + i} = \frac{1}{\mu} \, \frac{\displaystyle{\sum_{i = k - \alpha + 1}^k} \prod_{j = k - \alpha + 1 \atop j \neq i}^k j}{\displaystyle{\prod_{j = k - \alpha + 1}^k j}}.
\label{eq: Ts}
\end{equation}
Since $\muk = \frac{1}{\Ts}$,  (\ref{eq: muk}) is simply inferred from (\ref{eq: Ts}).
\end{IEEEproof}

While the $\muk$ expression in (\ref{eq: muk}) looks rather complicated, it is used in the following sections to simplify the service rate analysis in the form of the following corollary.

\begin{corol} \label{Corollary bound}
For $\alpha > 1$
\begin{align} \nonumber
\displaystyle{\sum_{i = k - \alpha + 1}^k} \prod_{j = k - \alpha + 1 \atop j \neq i}^k j & >  \displaystyle{\sum_{i = k - \alpha + 1}^k} \prod_{j = k - \alpha + 1 }^{k - 1} j \\  \label{eq: bound1}
& = \alpha \!\!\!  \prod_{j = k - \alpha + 1 }^{k - 1} j.
\end{align}
Now, using (\ref{eq: bound1}) and (\ref{eq: muk}) we have
\begin{equation}
\muk < \mu \frac{\displaystyle{\prod_{j = k - \alpha + 1 }^{k} j}}{\alpha \displaystyle{ \prod_{j = k - \alpha + 1 }^{k - 1} j}} =  \mu \frac{k}{\alpha}.
\label{eq: bound}
\end{equation}

\end{corol}

\section{Optimal Storage Allocation}
\subsection{Fixed-Size Access Model}\label{Sec Fixed}
In this section, we find the optimal allocation to maximize $\mumax$ for fixed-size access model.  For this, we start by the following lemma. 

\begin{lem}\label{Lemma mu1}
For minimal spreading, i.e. $\alpha = 1$, service rate is 
\begin{equation}
\mu_{\mathrm{s}}(1) = \mu \frac{mr}{N}.
\label{eq:}
\end{equation}
\end{lem}

\begin{IEEEproof}
First of all, note that for $\alpha = 1$, we have 
\begin{equation} \label{eq: muk1}
\mu_{\mathrm{s}}(1|k) = \mu k.
\end{equation}
Thus\footnote{Recall that for $k > m \alpha$, $\nchoosek{m \alpha}{k} = 0$, and hence, $P(k,\alpha) = 0$.}, 
\begin{equation}
\mu_{\mathrm{s}}(1) = \mu \sum_{k = 1}^{\min(m,r)} k P(k,1) = \mu \sum_{k = 1}^{r} k P(k,1).
\end{equation}
On the other hand, $\alpha = 1$ and
\begin{equation}
k \nchoosek{m \alpha}{k} = k \nchoosek{m }{k} = m \nchoosek{m  - 1}{k - 1}.
\end{equation}
As a result, 
\begin{equation}\label{eq: mu 1}
\mu_{\mathrm{s}}(1) = \frac{m \mu}{\nchoosek{N}{r}} \sum_{k = 1}^{r} \nchoosek{m - 1}{k - 1} \nchoosek{N - m}{r - k}.
\end{equation}
Using Vandermonde's convolution, one can show that 
\begin{equation}
\sum_{k = 1}^{r} \nchoosek{m - 1}{k - 1} \nchoosek{N - m}{r - k} = \nchoosek{N - 1}{r - 1}.
\label{eq: Vander}
\end{equation}
Now, plugging (\ref{eq: Vander}) into (\ref{eq: mu 1}) completes the proof.
\end{IEEEproof}

Now that we have $\mu_{\mathrm{s}}(1)$, the next step is to find an upper bound on $\mumax$ for any $2 \leq \alpha \leq r$ and compare this bound with $\mu_{\mathrm{s}}(1)$. First, we present the following lemma. 

\begin{lem} \label{Lemma bound}
For any $2 \leq \alpha \leq r$, 
\begin{equation}
\mumax < \mu \frac{mr}{N}.
\end{equation}
\end{lem}

\begin{IEEEproof}
Using Corollary~\ref{Corollary bound},
\begin{align}
\mumax & < \mu \sum_{k = \alpha}^{r} \frac{k}{\alpha} \Pk \\
& = \frac{\mu}{\alpha \nchoosek{N}{r}} \sum_{k = \alpha}^{r} k \nchoosek{m \alpha}{k} \nchoosek{N - m \alpha}{ r - k}\\ \label{eq: bound 2}
& = \frac{m \mu}{\nchoosek{N}{r}} \sum_{k = \alpha}^{r} \nchoosek{m \alpha - 1}{k - 1} \nchoosek{N - m \alpha}{r - k}.
\end{align}
In addition, 
\begin{align}\nonumber
\sum_{k = \alpha}^{r} \nchoosek{m \alpha - 1}{k - 1} \nchoosek{N - m \alpha}{r - k} & < \sum_{k = 0}^{r - 1} \nchoosek{m \alpha - 1}{k} \nchoosek{N - m \alpha}{r - 1 - k} \\ 
& = \nchoosek{N - 1}{r - 1}.
\end{align}
Hence,
\begin{equation}
\mumax < \frac{m \mu}{\nchoosek{N}{r}} \nchoosek{N - 1}{r - 1} = \mu \frac{mr}{N}.
\label{eq: mu final bound}
\end{equation}
\vspace{-0.05cm}
\end{IEEEproof}

%\begin{corol}
%$\mumax$ is a decreasing function of $\alpha$.
%\end{corol}

Now, using Lemma~\ref{Lemma mu1} and \ref{Lemma bound}, we have the following theorem on the optimal storage allocation maximizing $\mumax$.

\begin{theo} \label{Theorem fixed}
Minimal spreading maximizes the service rate for a DSS with fixed-size access model.
\end{theo}

\begin{figure}%
\includegraphics[width=\columnwidth]{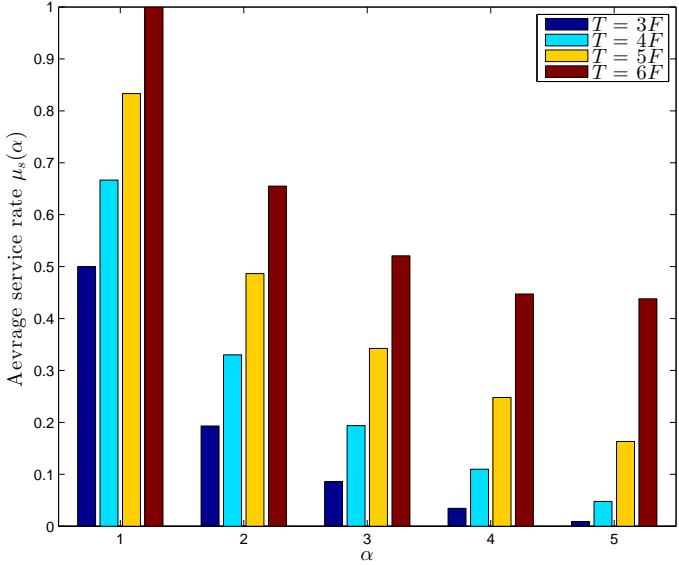}%
\caption{Average service rate for fixed-sized access model with $N = 30$ and $r = 5$.}%
\label{Fig: mu fixed}%
\end{figure}

\begin{figure}%
\includegraphics[width=.92\columnwidth]{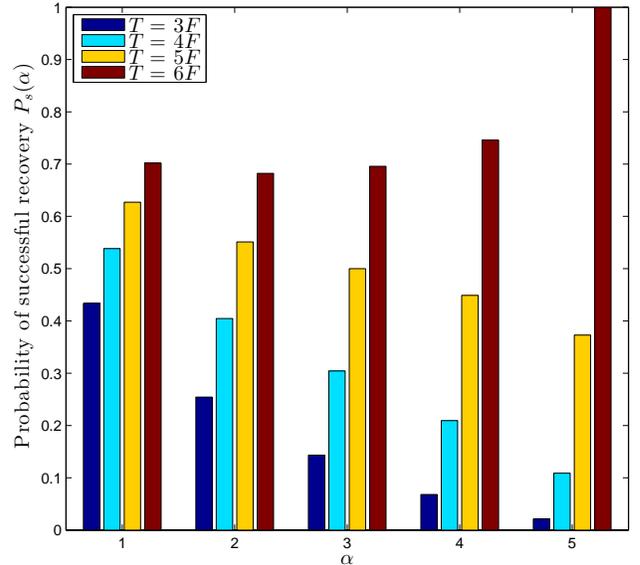}%
\caption{Probability of successful recovery for fixed-sized access model with $N = 30$ and $r = 5$.}%
\label{Fig: Ps fixed}%
\end{figure}

To verify the results of Theorem~\ref{Theorem fixed}, we present some numerical examples in Figure~\ref{Fig: mu fixed} and \ref{Fig: Ps fixed}. These figures depict the average service rate $\mumax$ and the probability of successful recovery $\Ps(\alpha)$ for a DSS with $N = 30$ nodes, a fixed-size access model with $r = 5$, and $\mu = 1$. Here, $\alpha = 1$ and $\alpha = 5$ are associated with minimal and maximal spreading allocation respectively. As seen in these figures, for smaller allocation budget $T$, minimal spreading maximizes both $\Ps(\alpha)$ and $\mumax$. However, as we increase $T$, while $\Ps(\alpha)$ is maximum when $\alpha = 5$ (in fact, any download request is successfully served and $\Ps(5) = 1$), $\mumax$ is always maximized by $\alpha = 1$.

\subsection{Probabilistic Access Model} \label{Sec Prob}
In this section, we study the service rate for a DSS with probabilistic access model. The goal is to find the optimal storage allocation maximizing the average service rate. 

\begin{lem} \label{Lem mu1 prob}
The service of a DSS with probabilistic access and minimal spreading allocation is 
\begin{equation}
\mu_{\mathrm{s}}(1) = m \mu (1 - p).
\label{eq: muk1 prob}
\end{equation}
\end{lem}

\begin{IEEEproof}
Using (\ref{eq: mumax def}), (\ref{eq:Pk prob}) and (\ref{eq: muk1}), we have
\begin{align}
\mu_{\mathrm{s}}(1) & = \mu  \sum_{k = 1}^{m} k \nchoosek{m}{k} (1 - p)^k p^{m - k} \\ 
& = m \mu \sum_{k = 1}^m \nchoosek{m - 1}{k - 1} (1 - p)^k p ^{m - k} \\
& = m \mu (1 - p) \sum_{k = 0}^{m - 1}\nchoosek{m - 1}{k}(1 - p)^k p^{m - k - 1} \\
& = m \mu (1 - p).
\end{align}
\end{IEEEproof}

Similar to the case of the fixed-size access model, we find an upper bound on the service rate of the system when $2 \leq \alpha$.

\begin{lem} \label{Lem mumax bound prob}
For a quasi-symmetric allocation where $2 \leq \alpha$, the service rate of the system is bounded as 
\begin{equation}
\mumax < m \mu (1 - p).
\label{eq: mus bound}
\end{equation}
\end{lem}

\begin{IEEEproof}
Using Corollary~\ref{Corollary bound}, we have
\begin{align}
\mumax &< \frac{\mu}{\alpha} \sum_{k = \alpha}^{m \alpha} k \nchoosek{m \alpha}{k} (1 - p)^k p^{m \alpha - k}\\ \nonumber
& = \frac{\mu}{\alpha} \sum_{k = \alpha}^{m \alpha} m \alpha \nchoosek{m \alpha - 1}{k - 1} (1 - p)^k p^{m \alpha - k}\\ \nonumber
& = m \mu (1 - p) \sum_{k = \alpha - 1}^{m \alpha - 1} \nchoosek{m \alpha - 1}{k} (1 - p)^k p^{m \alpha - k - 1}
\end{align}
On the other hand
\begin{equation}
\sum_{k = \alpha - 1}^{m \alpha - 1} \nchoosek{m \alpha - 1}{k} (1 - p)^k p^{m \alpha - k - 1} \leq 1.
\end{equation}
Thus,
\begin{equation}
\mumax < m \mu (1 - p).
\end{equation}
\vspace{-.05cm}
\end{IEEEproof}

Now, using the results of Lemma~\ref{Lem mu1 prob} and \ref{Lem mumax bound prob}, we have the following theorem on the optimal storage allocation for the probabilistic access model. 

\begin{theo}\label{Theo prob}
In a DSS with probabilistic access model, minimal spreading results in the maximum service rate. 
\end{theo}

\begin{figure}%
\includegraphics[width=\columnwidth]{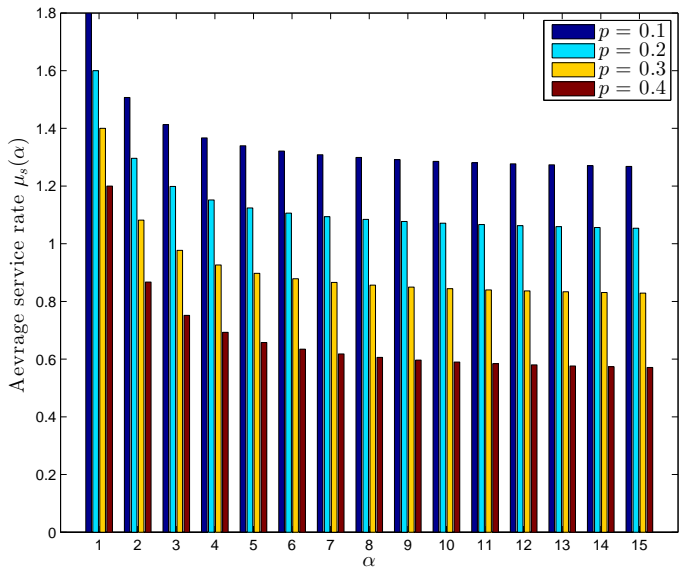}%
\caption{Average service rate for probabilistic access model when $N = 30$ and $T = 2F$.}%
\label{Fig: mu prob}%
\end{figure}

\begin{figure}%
\includegraphics[width=\columnwidth]{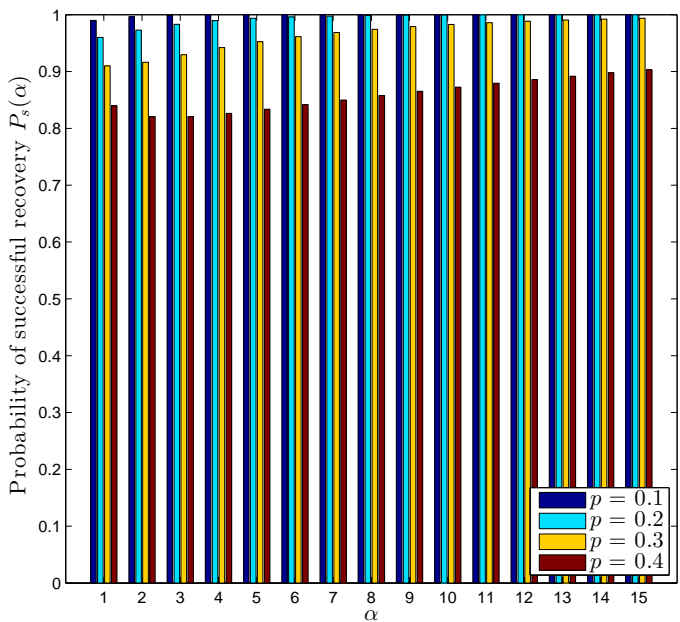}%
\caption{Probability of successful recovery for probabilistic access model when $N = 30$ and $T = 2F$.}%
\label{Fig: Ps prob}%
\end{figure} 

Numerical examples to verify the results of Theorem~\ref{Theo prob} are presented in Figure~\ref{Fig: mu prob} and \ref{Fig: Ps prob}. The results are for a DSS with $N = 30$ storage nodes, a storage budget of $T = 2F$, and $\mu = 1$. As seen in Figure~\ref{Fig: Ps prob}, maximal spreading allocation results in the highest probability of successful recovery for all considered probabilities of access failure. However, the average service rate always reaches its maximum for $\alpha = 1$, i.e. minimal spreading, as depicted in Figure~\ref{Fig: mu prob}.

\section{Conclusion} \label{Sec Conc}
Content allocation throughout a distributed storage system affects the probability that the content can be recovered when there is uncertainty in the number, identity, and/or availability of the storage nodes queried for service. So far the concern has been only that the stored data can eventually be downloaded, and not how long that process might take. To the best of our knowledge, this paper is the first attempt to understand how content allocation affects the download service rate.  We showed that under certain assumptions, the minimal spreading allocation maximizes the service rate for the commonly assumed content access models specified by the number, identity, and/or availability of the storage nodes queried for service.  Therefore, storing data through replication results in faster service for the incoming download requests than a coded storage with the same storage budget.  Our assumption was that the service time at the storage nodes follows an exponential distribution, and is identically distributed and independent for all users. A more advanced model should involve other distributions (in particular, the shifted exponential as in \cite{chen2014queueing} and \cite{joshi2012coding}) as well as fork-join queuing considerations. 

\section*{Acknowledgment}
The authors were in part supported by Alberta Innovates Technology Futures (AITF) and Natural Sciences and Engineering Research Council of Canada (NSERC), and would also like to thank 
A.~Badr, G.~Joshi, and K.~Mahdaviani for valuable discussions at the Banff International Research Station 
(BIRS).

\bibliographystyle{IEEEtran}
\bibliography{IEEEabrv,moslembib}

\end{document}